\RequirePackage{fix-cm}
\documentclass[smallextended]{svjour3}       
\smartqed  
\usepackage{graphicx}
\usepackage{amsmath}
\usepackage{subfigure}
\begin{document}

\title{Optimal shape of STIRAP pulses for large dissipation at the intermediate level
}
\subtitle{}


\author{Dionisis Stefanatos \and Emmanuel Paspalakis}


\institute{
              Materials Science Department, School of Natural Sciences, University of Patras, Patras 26504, Greece \\
              \email{dionisis@post.harvard.edu}           
}

\date{Received: date / Accepted: date}

\maketitle

\begin{abstract}
We study the problem of maximizing population transfer efficiency in the STIRAP system for the case where the dissipation rate of the intermediate state is much higher than the maximum amplitude of the control fields. Under this assumption, the original three-level system can be reduced to a couple of equations involving the initial and target states only. We find the control fields which maximize the population transfer to the target state for a given duration $T$, without using any penalty involving the population of the lossy intermediate state, but under the constraint that the sum of the intensities of the pump and Stokes pulses is constant, so the total field has constant amplitude and the only control parameter is the mixing angle of the two fields. In the optimal solution the mixing angle changes in the bang-singular-bang manner, where the initial and final bangs correspond to equal instantaneous rotations, while the intermediate singular arc to a linear change with time. We show that the optimal angle of the initial and final rotations is the unique solution of a transcendental equation where duration $T$ appears as a parameter, while the optimal slope of the intermediate linear change as well as the optimal transfer efficiency are expressed as functions of this optimal angle. The corresponding optimal solution recovers the counterintuitive pulse-sequence, with nonzero pump and Stokes fields at the boundaries. We also show with numerical simulations that, transfer efficiency values close to the optimal derived using the approximate system, can also be obtained with the original STIRAP system using dissipation rates comparable to the maximum control amplitude.
\end{abstract}

\section{Introduction}
\label{intro}

Stimulated Raman adiabatic passage (STIRAP) is a very efficient method for population transfer between two quantum states, say $|1\rangle$ and $|3\rangle$, which are indirectly connected through a lossy state $|2\rangle$ \cite{Bergmann98,Kral07,Vitanov17,Sola18,Bergmann19,Vitanov97,Kobrak98}. This intermediate state is coupled to the initial and target states with the pump and Stokes laser pulses, respectively. In the STIRAP pulse-sequence, the Stokes pulse precedes the pump pulse, in a counterintuitive order, building a coherent superposition of states $|1\rangle$ and $|3\rangle$ which, in the course of time, evolves from the starting state to the target state. If the change in the control fields is slow enough, the intermediate lossy state is hardly populated, and complete population transfer is achieved. STIRAP has proven to be quite robust against moderate variations of experimental parameters and has been applied to a wide spectrum of modern physical systems \cite{Bergmann19}, including optical waveguides \cite{Paspalakis06,Dreisow09}, matter waves \cite{Menchon16}, nitrogen-vacancy centers in diamond \cite{Golter14}, and superconducting quantum circuits \cite{Kumar16}.

In order to improve the performance of STIRAP, optimal control \cite{Bryson} has been employed. Initially, it was proven that the counterintuitive pulse-sequence with pulses starting from zero, cannot be the solution of an optimal control problem with objective the maximization of the population of the target state \cite{Band94}. Nevertheless, it was soon demonstrated using numerical optimal control that, if a penalty term which punishes populating the lossy intermediate level is added in the cost function, then the counterintuitive pulse-sequence is recovered as the optimal solution \cite{Sola99,Kis02,Kumar11}. In a more theoretical work \cite{Boscain02}, where the dissipation term was omitted and the goal was to minimize the transfer time or the fluence of the control fields, the optimal solution was not the STIRAP counterintuitive sequence but the intuitive one, with the pump pulse preceding the Stokes pulse. In another theoretical work \cite{Rat12}, which took into account dissipation, using the techniques introduced in \cite{Khaneja03,Stefanatos04,Stefanatos05} and employing bounded Stokes and unbounded pump pulses, an optimal solution mimicking the counterintuitive pulse-sequence was derived. The corresponding optimal Stokes pulse was constant while the pump pulse was much smaller at the beginning and much larger at the end. In Ref. \cite{Assemat12}, which also considered dissipation, an optimal counterintuitive pulse-sequence was obtained with nonzero Stokes and pump pulses at the initial and final times, respectively, using again a cost function penalizing the occupation of the lossy intermediate state. Finally, in Ref. \cite{Dalessandro20} dissipation is ignored and the minimum transfer time optimal control problem is considered. Besides the above works, which use analytical or numerical optimal control methods to maximize STIRAP efficiency, several other methods, belonging in the family of shortcuts to adiabaticity \cite{STA19}, have been developed \cite{Demirplak05,Chen10b,Giannelli14,Masuda15,Li16,Clerk16,Kolbl19,Dridi20,Petiziol20}.

In the present work we study the STIRAP problem in the limit where the dissipation rate of the intermediate level is much higher than the maximum amplitude of the control fields used. This condition allows us to adiabatically eliminate from STIRAP equations the probability amplitude of state $|2\rangle$, reducing thus the original three-level system to two coupled equations involving the probability amplitudes of states $|1\rangle$ and $|3\rangle$. In this reduced system we find the control fields which maximize the population transfer from state $|1\rangle$ to state $|3\rangle$ for a given duration $T$, without using any penalty involving the population of state $|2\rangle$. In order to solve this optimal control problem, we take the sum of the intensities of the pump and Stokes pulses to be constant, so the total field has constant amplitude and the only control parameter is the mixing angle of the two fields. In the optimal solution that we derive, the mixing angle changes in the bang-singular-bang manner, where the initial and final bangs correspond to instantaneous rotations of equal amounts, while the intermediate singular arc to a linear change with time. The boundary rotations and the slope of the intermediate linear variation depend on $T$ and decrease as the duration increases. Specifically, the optimal angle of the initial and final instantaneous rotations is the unique solution of a transcendental equation where $T$ appears as a parameter, while the optimal slope as well as the optimal efficiency are expressed as functions of this optimal angle. The corresponding optimal solution recovers the counterintuitive pulse-sequence, with nonzero pump and Stokes fields at the boundaries, in consistency with the results of Ref. \cite{Band94}. We also demonstrate with numerical simulations that, transfer efficiency values close to the optimal derived using the approximate system, can also be obtained with the original STIRAP system using dissipation rates comparable to the maximum control amplitude.

\section{Formulation of the problem}
\label{sec:formulation}

The time evolution of probability amplitudes $c_i, i=1,2,3$, for the STIRAP system, in both one-photon and two-photon resonance, is governed by the following equation
\begin{equation}
\label{STIRAP}
i
\left (
\begin{array}{c}
\dot{c}_1 \\
\dot{c}_2\\
\dot{c}_3
\end{array}
\right )
=
\frac{1}{2}
\left(
\begin{array}{ccc}
0 & \Omega_p(t)  & 0\\
\Omega_p(t) & -i\Gamma & \Omega_s(t) \\
0 & \Omega_s(t)  & 0
\end{array}
\right)
\left (
\begin{array}{c}
c_1 \\
c_2\\
c_3
\end{array}
\right )
\end{equation}
where $\Omega_p(t), \Omega_s(t)$ are the Rabi frequencies of the pump and Stokes lasers, respectively, and $\Gamma$ is the population dissipation rate from level $|2\rangle$. The initial conditions at $t=0$ are
\begin{equation}
\label{initial_conditions}
c_1(0)=1, \quad c_2(0)=c_3(0)=0,
\end{equation}
while the goal is to find the controls fields $\Omega_p(t), \Omega_s(t)$ which maximize the population of level $|3\rangle$, $|c_3(T)|^2$, at a given final time $t=T$.

In order to make this control problem well defined, we impose a bound on the control amplitudes through the constraint
\begin{equation}
\label{bounded_fluence}
\Omega^2_p(t)+\Omega^2_s(t)=\Omega^2_0\;\mbox{constant},
\end{equation}
corresponding to constant magnitude of the total field in the equivalent two-level picture. This choice leaves as control variable the direction of the total field, determined by the
mixing angle
\begin{equation}
\label{mixing_angle}
\tan{\theta(t)}=\frac{\Omega_p(t)}{\Omega_s(t)}.
\end{equation}
Furthermore, we consider the case where dissipation is large compared to the fields, $\Omega_0\ll\Gamma$. Under this condition, we can eliminate adiabatically $c_2$ from Eg. (\ref{STIRAP}) using the relation
\begin{equation}
\label{ad_elimination}
c_2=-\frac{i}{\Gamma}(\Omega_pc_1+\Omega_sc_3),
\end{equation}
which is obtained by setting $\dot{c}_2=0$, and find the following system of equations for $c_1, c_3$ only \cite{Vitanov97}
\begin{equation}
\label{eliminated}
\left (
\begin{array}{c}
\dot{c}_1 \\
\dot{c}_3
\end{array}
\right )
=
-\frac{1}{2\Gamma}
\left(
\begin{array}{cc}
\Omega^2_p  & \Omega_p\Omega_s\\
\Omega_p\Omega_s &  \Omega^2_s
\end{array}
\right)
\left (
\begin{array}{c}
c_1 \\
c_3
\end{array}
\right ).
\end{equation}
Using Eqs. (\ref{bounded_fluence}), (\ref{mixing_angle}) and normalizing time as $\Omega_0^2t/\Gamma$, we obtain
\begin{equation}
\label{original}
\left (
\begin{array}{c}
\dot{c}_1 \\
\dot{c}_3
\end{array}
\right )
=
-\frac{1}{2}
\left(
\begin{array}{cc}
\sin^2\theta  & \sin\theta\cos\theta\\
\sin\theta\cos\theta & \cos^2\theta
\end{array}
\right)
\left (
\begin{array}{c}
c_1 \\
c_3
\end{array}
\right ).
\end{equation}

In order to study the system evolution under a time-dependent mixing angle $\theta(t)$, it is more convenient to work in the adiabatic frame. The eigenstates of system (\ref{original}) are
\begin{equation}
\label{eigenstates}
\psi_{0}=\left(
\begin{array}{c}
  \cos{\theta}\\
  -\sin{\theta}
\end{array}
\right)
,
\quad
\psi_{-1/2}=\left(
\begin{array}{c}
  \sin{\theta}\\
  \cos{\theta}
\end{array}
\right),
\end{equation}
with corresponding eigenvalues $0$ and $-1/2$, respectively. If we express an arbitrary state of system (\ref{original}) in both bases, $(c_1, c_3)^T=y\psi_{0}+x\psi_{-1/2}$, we find
\begin{equation}
\label{direct}
\left(
\begin{array}{c}
  c_1\\
  c_3
\end{array}
\right)
=
\left(
\begin{array}{cc}
  \cos{\theta} & \sin{\theta}\\
  -\sin{\theta} & \cos{\theta}
\end{array}
\right)
\left(
\begin{array}{c}
  y\\
  x
\end{array}
\right),
\end{equation}
thus the inverse transformation to the adiabatic basis is
\begin{equation}
\label{transformation}
\left(
\begin{array}{c}
  y\\
  x
\end{array}
\right)
=
\left(
\begin{array}{cc}
  \cos{\theta} & -\sin{\theta}\\
  \sin{\theta} & \cos{\theta}
\end{array}
\right)
\left(
\begin{array}{c}
  c_1\\
  c_3
\end{array}
\right).
\end{equation}
Using Eqs. (\ref{original}), (\ref{direct}) and (\ref{transformation}) we find the system equation in the adiabatic basis
\begin{equation}
\label{adiabatic}
\left(
\begin{array}{c}
  \dot{y}\\
  \dot{x}
\end{array}
\right)
=
\left(
\begin{array}{cc}
  0 & -u\\
  u & -\frac{1}{2}
\end{array}
\right)
\left(
\begin{array}{c}
  y\\
  x
\end{array}
\right),
\end{equation}
where the control function $u(t)$ is defined as
\begin{equation}
\label{theta}
\dot{\theta}=u(t).
\end{equation}
Imposing on angle $\theta$ the boundary conditions
\begin{equation}
\label{b_theta}
\theta(0)=0,\quad\theta(T)=\frac{\pi}{2}
\end{equation}
we get from Eq. (\ref{adiabatic})
\begin{equation}
\label{initial}
x(0)=c_3(0)=0,\quad y(0)=c_1(0)=1
\end{equation}
and
\begin{equation}
\label{final}
x(T)=c_1(T),\quad y(T)=c_3(T).
\end{equation}

Note that, for large durations $T$, if the angle $\theta$ is slowly increased from $0$ to $\pi/2$, with rate $u=\dot{\theta}\ll 1$, then $y(t)$ remains constant during the evolution, thus $c_3(T)=y(T)\approx y(0)=1$. This slow transfer between states $|1\rangle$ and $|3\rangle$ takes place along the dark eigenstate $\psi_0$ of the original system (\ref{original}).
In the next section we find the optimal control $u(t)$, $0\leq t\leq T$, for given finite duration $T$, which maximizes the final value $|c_3(T)|^2=y^2(T)$. Note that $u=\dot{\theta}$ corresponds to the adiabatic gauge potential which represents nonadiabaticity, and optimal control theory allows us to determine how this control function should be varied in order to maximize the desired objective. The mathematical problem is similar to that of maximizing the conversion efficiency between light beams of different frequency or orbital angular momentum, propagating in a cloud of cold atoms characterized by a double-$\Lambda$ atom-light coupling scheme \cite{Stefanatos20f}. We present the detailed solution below.

\section{Optimal solution}
\label{optimal_sol}

In order to solve the optimal control problem defined in the previous section, we need to formulate the control Hamiltonian \cite{Bryson} corresponding to system (\ref{adiabatic}), (\ref{theta}). This is a mathematical construction whose maximization results in the maximization of the target quantity, here $y(T)$. It is formed by adjoining to each state equation a conjugate variable (Lagrange multiplier) as follows
\begin{equation}
\label{Hc}
H_c=\lambda_x\dot{x}+\lambda_y\dot{y}+\mu u=(\lambda_x y-\lambda_y x+\mu)u-\frac{1}{2}\lambda_x x,
\end{equation}
where $\lambda_x,\lambda_y,\mu$ are the Lagrange multipliers corresponding to state variables $x,y,\theta$. They satisfy the adjoint equations
\begin{subequations}
\label{lambda}
\begin{eqnarray}
\dot{\lambda}_y&=-\frac{\partial H_c}{\partial y}&=-u\lambda_x,\label{ly}\\
\dot{\lambda}_x&=-\frac{\partial H_c}{\partial x}&=u\lambda_y+\frac{1}{2}\lambda_x,\label{lx}
\end{eqnarray}
\end{subequations}
while $\mu$ is constant since $\theta$ is a cyclic variable. Due to the construction of $H_c$ the state equations can also be expressed as
\begin{subequations}
\begin{eqnarray}
\dot{y}&=\frac{\partial H_c}{\partial \lambda_y}&=-ux,\\
\dot{x}&=\frac{\partial H_c}{\partial \lambda_x}&=uy-\frac{1}{2}x,\\
\dot{\theta}&=\frac{\partial H_c}{\partial \mu}&=u,
\end{eqnarray}
\end{subequations}
which justifies the term Hamiltonian for $H_c$. 

According to optimal control theory \cite{Bryson}, the control function $u(t)$ is chosen to maximize the control Hamiltonian $H_c$. Note that we have not imposed any bound on $u$, so Eqs. (\ref{adiabatic}) and (\ref{theta}) are fully equivalent to the original system (\ref{original}). Even infinite values are allowed momentarily, corresponding to instantaneous jumps in the angle $\theta$. Since $H_c$ is a linear function of $u$ with coefficient $\phi=\lambda_x y-\lambda_y x+\mu$, if $\phi\neq 0$ for a finite time-interval then the corresponding optimal control should be $\pm\infty$ for the whole interval, which is obviously unphysical. We conclude that $\phi=0$ almost everywhere, except some isolated points where jumps in the angle $\theta$ can occur. The optimal control which maintains this condition is called \emph{singular} \cite{Bryson}. Such controls have been exploited in nuclear magnetic resonance to minimize the effect of relaxation \cite{Lapert10,Lin20}, but we have also used singular control in the problem of maximizing entanglement between two quantum oscillators with time-dependent coupling \cite{Stefanatos17a}. In order to find the singular optimal control $u_s$ we additionally use the conditions $\dot{\phi}=\ddot{\phi}=0$ and obtain the following equations
\begin{subequations}
\label{phi}
\begin{eqnarray}
\lambda_y x-\lambda_x y&=&\mu,\label{p}\\
\lambda_y x+\lambda_x y&=&0,\label{dp}\\
2(\lambda_y y-\lambda_x x)u_s&=&\frac{1}{2}(\lambda_y x-\lambda_x y).\label{ddp}
\end{eqnarray}
\end{subequations}
Solving for $\lambda_x,\lambda_y,u_s$ we find
\begin{equation}
\label{lxy}
\lambda_x=-\frac{\mu}{2y},\quad \lambda_y=\frac{\mu}{2x},
\end{equation}
and
\begin{equation}
\label{feedback}
u_s=\frac{xy}{2(x^2+y^2)}.
\end{equation}

According to optimal control theory \cite{Bryson}, the control Hamiltonian for a system without explicit time dependence, as in our case, is constant. Using $\phi=0$ and the expression (\ref{lxy}) for $\lambda_x$  in Eq. (\ref{Hc}), we conclude that the singular arc is a straight line passing through the origin of the $xy$-plane,
\begin{equation}
\label{slope}
\frac{x}{y}=\mbox{constant}=\tan{\theta_0}.
\end{equation}
This implies that the singular control $u_s$, given in Eq. (\ref{feedback}) in terms of a feedback law, is also constant. Now we can describe the optimal pulse-sequence. There is a delta pulse at $t=0$, resulting in a jump from $\theta(0^-)=0$ to $\theta(0^+)=\theta_0$, where the initial angle is to be determined. The system is brought on the singular arc and remains there for $0<t<T$. During this interval, $\theta$ increases linearly with slope $u_s$, $\theta(t)=\theta_0+u_st$, following Eq. (\ref{theta}) with constant $u(t)=u_s$. At the final time $t=T$, another delta pulse changes $\theta$ from $\theta(T^-)=\theta_0+u_sT$ to the final value $\theta(T^+)=\pi/2$. The optimal control has the form \emph{bang-singular-bang},
\begin{equation}
\label{pulse_sequence}
u(t)=\left\{\begin{array}{cl} \theta_0\delta(t), & t=0 \\u_s, & 0<t<T\\ (\pi/2-\theta_0-u_sT)\delta(t-T), & t=T \end{array}\right..
\end{equation}
Observe from the original system (\ref{original}) that jumps in the angle $\theta$ do not change $c_1,c_3$. In the adiabatic basis these jumps are accompanied by sudden rotations of the $(x,y)^T$ vector such that $c_1,c_3$ remain unchanged, see the transformation (\ref{direct}). Thus, in order to implement the optimal protocol, one simply needs to vary linearly the angle from $\theta_0$ to $\theta_0+u_sT$ with the slope $u_s<\pi/(2T)$. 

We now find the optimal $\theta_0$ corresponding to a given duration $T$. For this purpose we need to express the final value $|c_3(T)|^2=y^2(T)$ in terms of $\theta_0$, and then optimize with respect to this variable.  After the application of the first delta pulse to the adiabatic system (\ref{adiabatic}), the initial state $(x(0^-),y(0^-))^T=(0,1)^T$ is rotated clockwise by an angle $\theta_0$ to the state $(x(0^+),y(0^+))^T=(\sin\theta_0,\cos\theta_0)^T$. The constant singular control $u_s$ can be easily expressed in terms of $\theta_0$ using Eqs. (\ref{feedback}), (\ref{slope}), and the result is
\begin{equation}
\label{u_theta}
u_s=\frac{\tan{\theta_0}}{2(1+\tan^2{\theta_0})}=\frac{1}{4}\sin{(2\theta_0)}.
\end{equation}
Using Eq. (\ref{slope}) in Eq. (\ref{adiabatic}), it can be easily shown that on the singular arc the adiabatic states evolve according to the equations $\dot{y}=-\gamma y, \dot{x}=-\gamma x$, where
\begin{equation}
\label{gamma}
\gamma=u_s\tan{\theta_0}=\frac{1}{4}[1-\cos{(2\theta_0)}],
\end{equation}
thus $(x(T^-),y(T^-))^T=e^{-\gamma T}(\sin\theta_0,\cos\theta_0)^T$. The final delta pulse rotates the vector of adiabatic states clockwise by an angle $\pi/2-\theta_0-u_sT$, thus $y(T^+)=e^{-\gamma T}\cos(\pi/2-u_sT)=e^{-\gamma T}\sin(u_sT)$ and
\begin{equation}
\label{max_eff}
|c_3(T)|^2=y^2(T)=e^{-2\gamma T}\sin^2{(u_sT)}.
\end{equation}

If we optimize the right hand side of Eq. (\ref{max_eff}) with respect to $\theta_0$ we obtain
\begin{equation}
\label{preliminary}
\tan{(u_sT)}=\frac{du_s}{d\theta_0}/\frac{d\gamma}{d\theta_0}=\cot(2\theta_0)=\tan{\left(\frac{\pi}{2}-2\theta_0\right)}.
\end{equation}
Since $0<u_sT<\pi/2$ and $0<\theta_0<\pi/4$, where the upper bound in the latter relation comes from the third term in Eq. (\ref{preliminary}), from the monotonicity of the $\tan$ function in $(0,\pi/2)$ we conclude that
\begin{equation}
\label{transcendental}
\frac{T}{4}\sin{(2\theta_0)}=\frac{\pi}{2}-2\theta_0,
\end{equation}
where we have also used Eq. (\ref{u_theta}). For a given duration $T$, this is a transcendental equation for $\theta_0$. It can be easily proved that it has a unique solution in the interval $0<\theta_0<\pi/4$. Using Eqs. (\ref{u_theta}), (\ref{gamma}) and (\ref{transcendental}) in Eq. (\ref{max_eff}), we can express the optimal transfer efficiency in terms of the optimal $\theta_0$,
\begin{equation}
\label{opt_eff}
|c_3(T)|^2=e^{-\tan{\theta_0}(\pi-4\theta_0)}\cos^2{(2\theta_0)}.
\end{equation}
Note that for $T\rightarrow 0$ it is $\theta_0\rightarrow\pi/4$, while for $T\rightarrow \infty$ it is $\theta_0\rightarrow 0$. Using Eqs. (\ref{transcendental}) and (\ref{opt_eff}) we find the limiting values
\begin{equation}
\label{limits_opt}
|c_3(T)|^2=\left\{\begin{array}{cl} \frac{1}{16}T^2, & T\ll 1 \\1-\frac{\pi^2}{T}, & T\gg 1\end{array}\right.,
\end{equation}
thus for sufficiently large durations the transfer efficiency approaches unity.

\begin{figure}[t]
\centering
\includegraphics[width=0.6\linewidth]{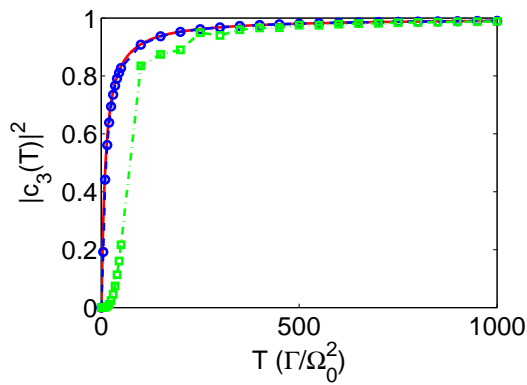}
\caption{Optimal transfer efficiency (red solid line) as a function of normalized duration. The transfer efficiencies obtained when the optimal pulse-sequence is applied in the original system (\ref{STIRAP}) with $\Gamma/\Omega_0=1$ (blue circles) and $\Gamma/\Omega_0=0.1$ (green squares) are also shown. }
\label{fig:efficiency}
\end{figure}

In Fig. \ref{fig:efficiency} we plot the optimal efficiency as a function of the normalized duration, for $0\leq T\leq 1000$. We also plot the transfer efficiency obtained when the optimal pulse-sequence is applied in the original system (\ref{STIRAP}) with $\Gamma/\Omega_0=1$ (blue circles) and $\Gamma/\Omega_0=0.1$ (green squares), using a step $\delta T=5$ for $0\leq T\leq 50$ and $\delta T=50$ for $50\leq T\leq 1000$. Observe that for $\Gamma/\Omega_0=1$ the achieved transfer efficiency is close to the optimal, while for $\Gamma/\Omega_0=0.1$ the optimal is approached only for larger normalized durations. This behavior can be understood if for the optimal pulse-sequence we take $\dot{\theta}\approx \pi/(2T)\times\Omega^2_0/\Gamma$, where note that $T$ is the normalized duration in units of $\Gamma/\Omega_0^2$, and use it in the adiabaticity condition for the original system $\dot{\theta}\ll \Omega_0/2$ (neglecting dissipation), which lead to the condition
\begin{equation}
\label{adiabaticity_optimal}
\frac{\pi}{T}\ll \frac{\Gamma}{\Omega_0}.
\end{equation}
For $\Gamma/\Omega_0=0.1$ this relation requires $T\gg 30$. Another way to understand Fig. \ref{fig:efficiency} is that, since $T$ denotes the normalized duration in units of $\Gamma/\Omega^2_0$, for smaller $\Gamma/\Omega_0$ the same normalized duration corresponds to less actual duration and thus to a lower efficiency obtained with the original system.

As we pointed out above, for large $T$ we have $\theta_0\rightarrow 0$, thus the boundary jumps become smaller. In this limit, the optimal solution tends to a linear increase of the angle $\theta$ from $0$ to $\pi/2$,
\begin{equation}
\label{constant_u}
u(t)=\mbox{const.}=\frac{\pi}{2T},\quad\theta(t)=ut.
\end{equation}
Under this constant control protocol, ensuring that $\Omega_s(0)=\Omega_p(T)=\Omega_0$ and $\Omega_s(T)=\Omega_p(0)=0$, Eq.\ (\ref{adiabatic}) can be easily integrated and at the final time $t=T$ one finds
\begin{equation}
\label{constant_eff}
|c_3(T)|^2=e^{-\eta T}\left[\cosh{(\kappa T)}+\frac{\eta}{2\kappa}\sinh{(\kappa T)}\right]^2,
\end{equation}
where
\begin{equation*}
\eta=\frac{1}{2},\quad\kappa=\sqrt{\left(\frac{\eta}{2}\right)^2-u^2}.
\end{equation*}
From Eq. (\ref{constant_eff}) we find the limiting cases
\begin{equation}
\label{limits_const}
|c_3(T)|^2=\left\{\begin{array}{cl} \frac{1}{4\pi^2}T^2, & T\ll 1 \\1-\frac{\pi^2}{T}, & T\gg 1\end{array}\right..
\end{equation}
Compared to the corresponding values (\ref{limits_opt}) of the optimal protocol, the constant control protocol (\ref{constant_u}) behaves worse for smaller values of $T$, while both strategies behave similarly for large $T$. 

\section{Example and discussion}

\begin{figure}[t]
 \centering
		\begin{tabular}{cc}
       \subfigure[Optimal control fields]{
	            \label{fig:opt_controls}
	            \includegraphics[width=0.5\linewidth]{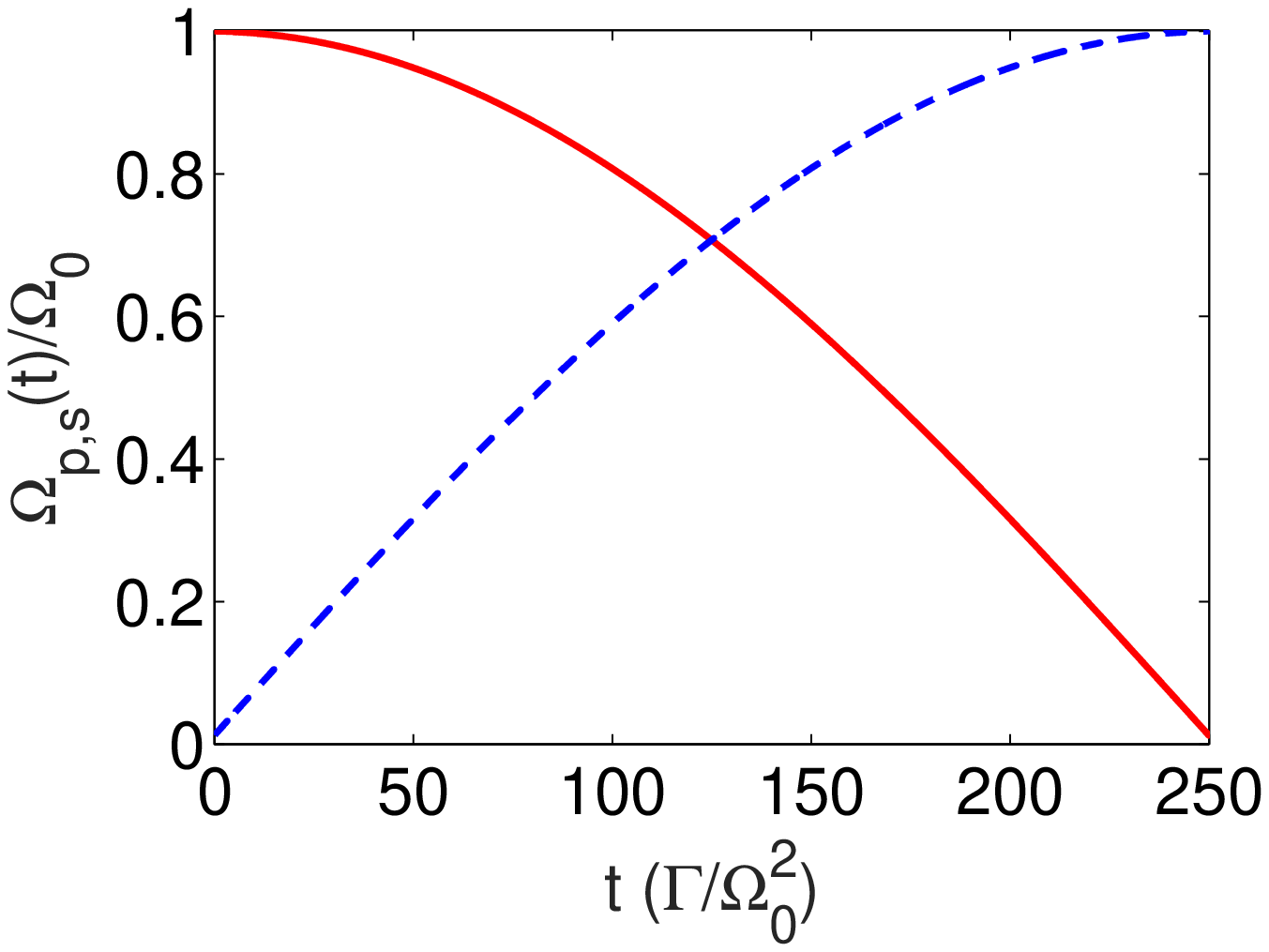}} &
       \subfigure[Adiabatic control fields]{
	            \label{fig:ad_controls}
	            \includegraphics[width=0.5\linewidth]{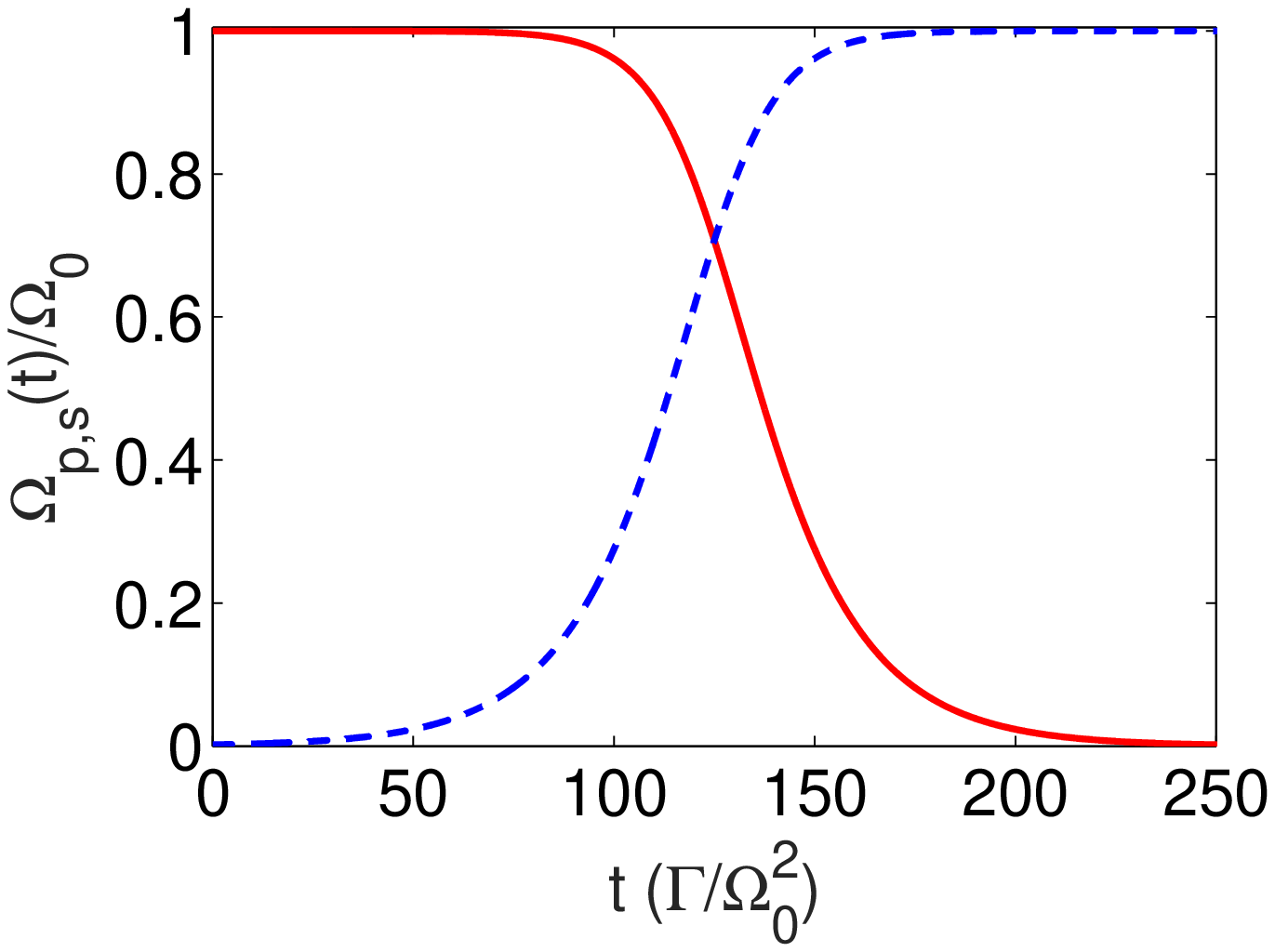}}  \\
       \subfigure[Populations for the optimal pulses]{
	            \label{fig:opt_pop}
	            \includegraphics[width=0.5\linewidth]{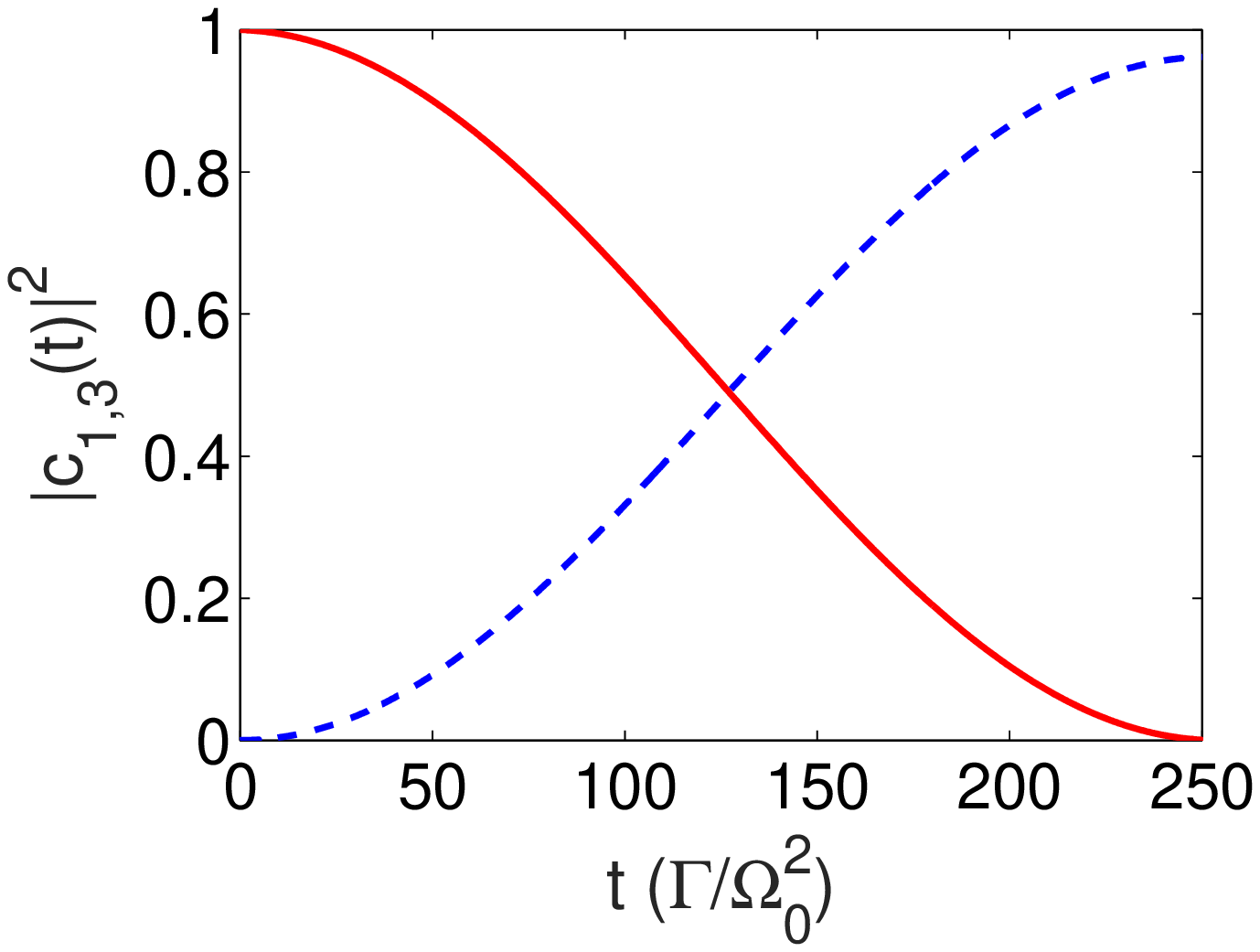}} &
       \subfigure[Populations for the adiabatic pulses]{
	            \label{fig:ad_pop}
	            \includegraphics[width=0.5\linewidth]{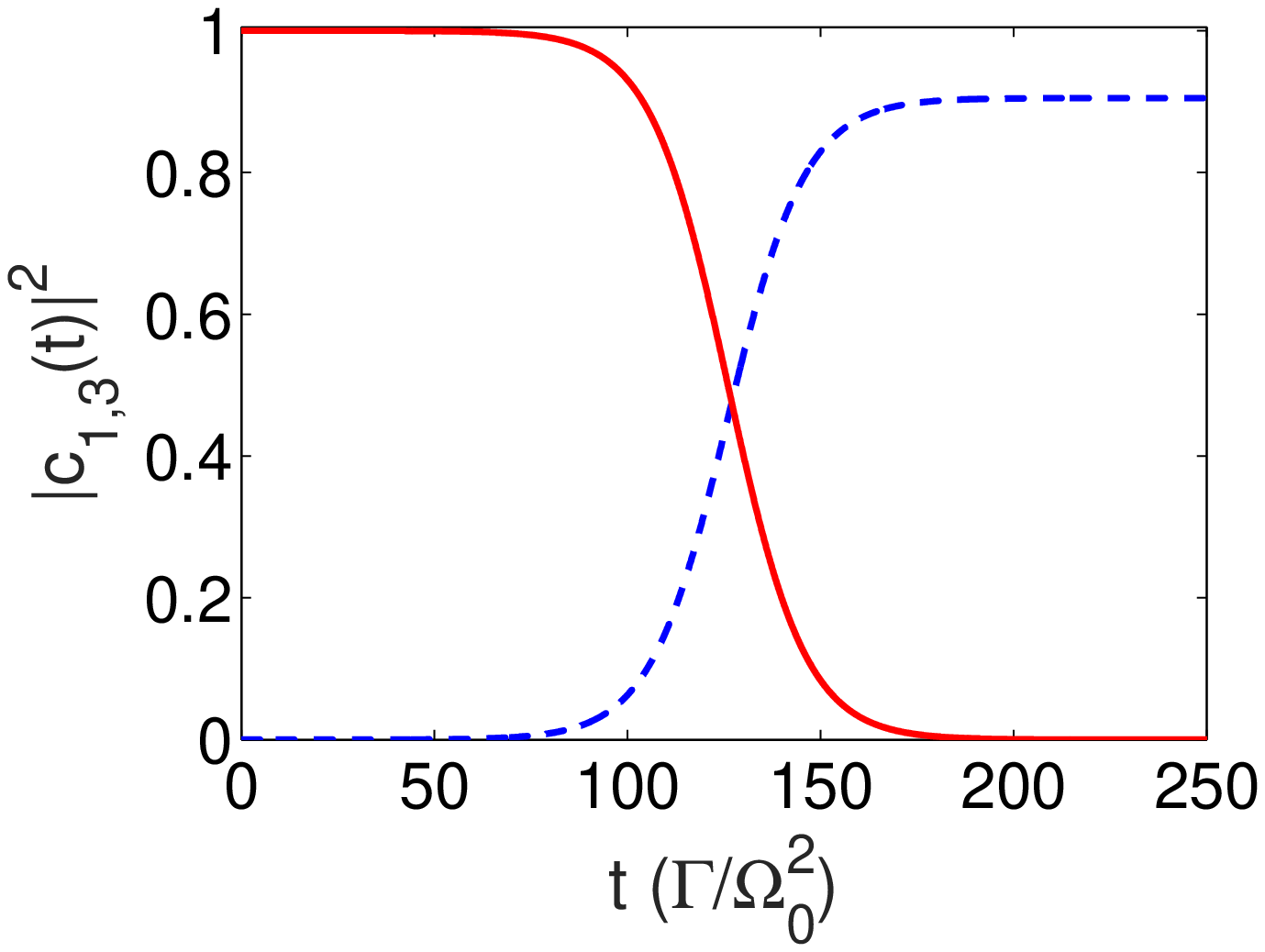}}
		\end{tabular}
\caption{(a) Optimal pump (blue dashed line) and Stokes (red solid line) pulses for normalized duration $T=250$. (b) Adiabatic pulses of Eq. (\ref{adiabatic_controls}).
(c) Evolution of populations of levels $|1\rangle$ (red solid line) and $|3\rangle$ (blue dashed line) for the optimal pulses. (d) Evolution of populations for the adiabatic pulses.}
\label{fig:example}
\end{figure}

In this section we consider a concrete example for normalized duration $T=250$. The numerical solution of the corresponding transcendental Eq.\ (\ref{transcendental}) gives the optimal value $\theta_0\approx 0.012370$ rad, while from Eq. (\ref{u_theta}) we obtain $u_s\approx 0.006184$. The corresponding optimal pump and Stokes pulses are displayed in Fig. \ref{fig:opt_controls}. In general, the Stokes field $\Omega_s$ precedes the pump field $\Omega_p$ in a counterintuitive sequence, to prepare the necessary coherence for the transfer $|1\rangle\rightarrow |3\rangle$, but note that for the optimal protocol $\Omega_p(0^+)$ and $\Omega_s(T^-)$ have nonzero values, associated with the jumps in the mixing angle at $t=0$ and $t=T$, respectively. The corresponding populations $|c_1(t)|^2, |c_3(t)|^2$ are plotted in Fig. \ref{fig:opt_pop}, and the achieved transfer efficiency is $|c_3(T)|^2=0.962$.
For comparison, we consider a pair of adiabatic controls that we have used in Refs. \cite{Paspalakis02,Hamedi19}, of the form
\begin{equation}
\label{adiabatic_controls}
\Omega_s=\Omega_0\left[1+e^{(t-t_0)/\bar{T}}\right]^{-1/2},\quad \Omega_p=\Omega_0\left[1+e^{-(t-t_0)/\bar{T}}\right]^{-1/2}.
\end{equation}
Here, we take $t_0=T/2=125$ and $\bar{T}=10$. The control fields are shown in Fig. \ref{fig:ad_controls}, while the corresponding populations in Fig. \ref{fig:ad_pop}. The transfer efficiency is now $|c_3(T)|^2=0.905$, lower than the value obtained with the optimal protocol.

\begin{figure}[t]
 \centering
		\begin{tabular}{cc}
       \subfigure[Optimal control fields]{
	            \label{fig:opt_finite}
	            \includegraphics[width=0.5\linewidth]{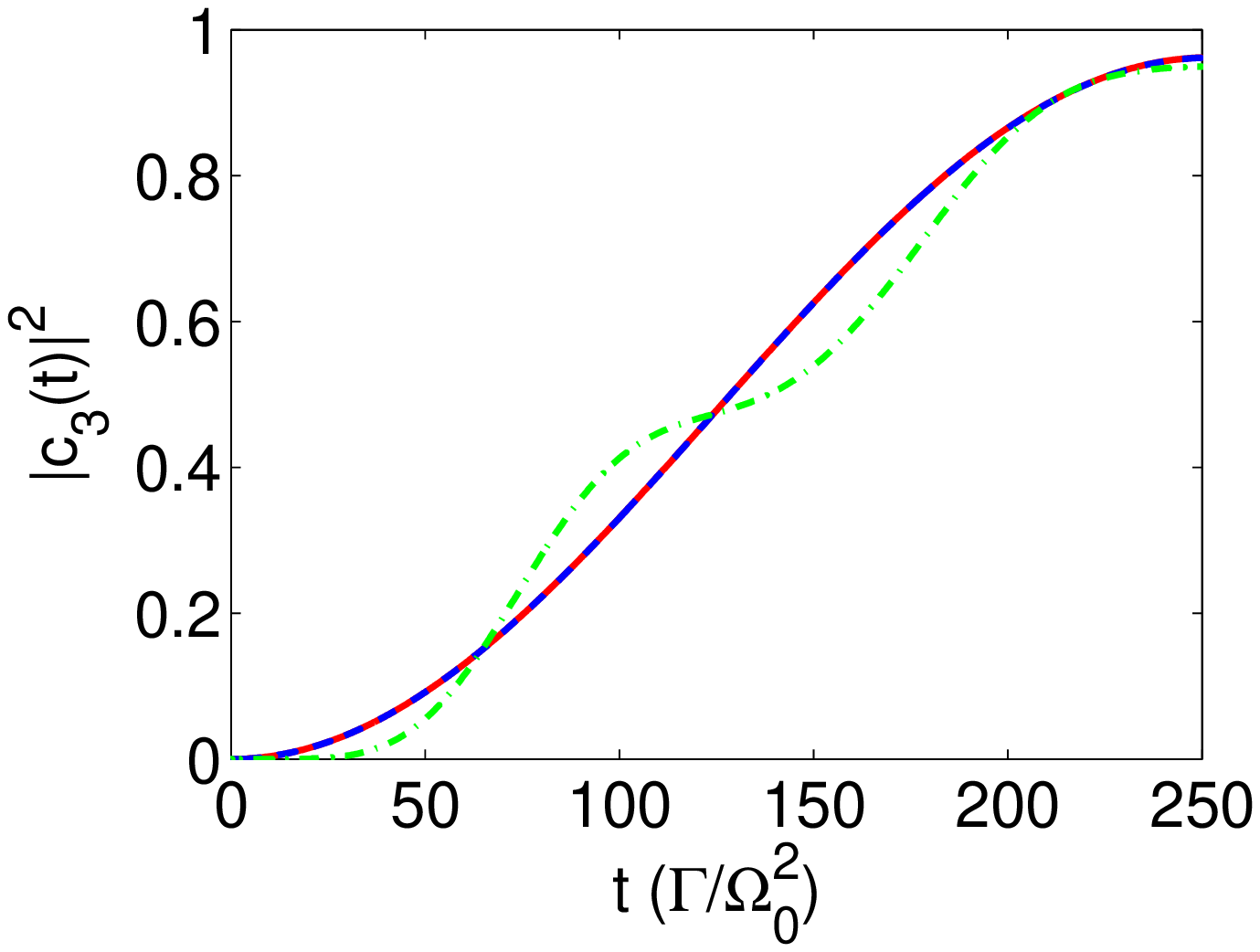}} &
       \subfigure[Adiabatic control fields]{
	            \label{fig:ad_finite}
	            \includegraphics[width=0.5\linewidth]{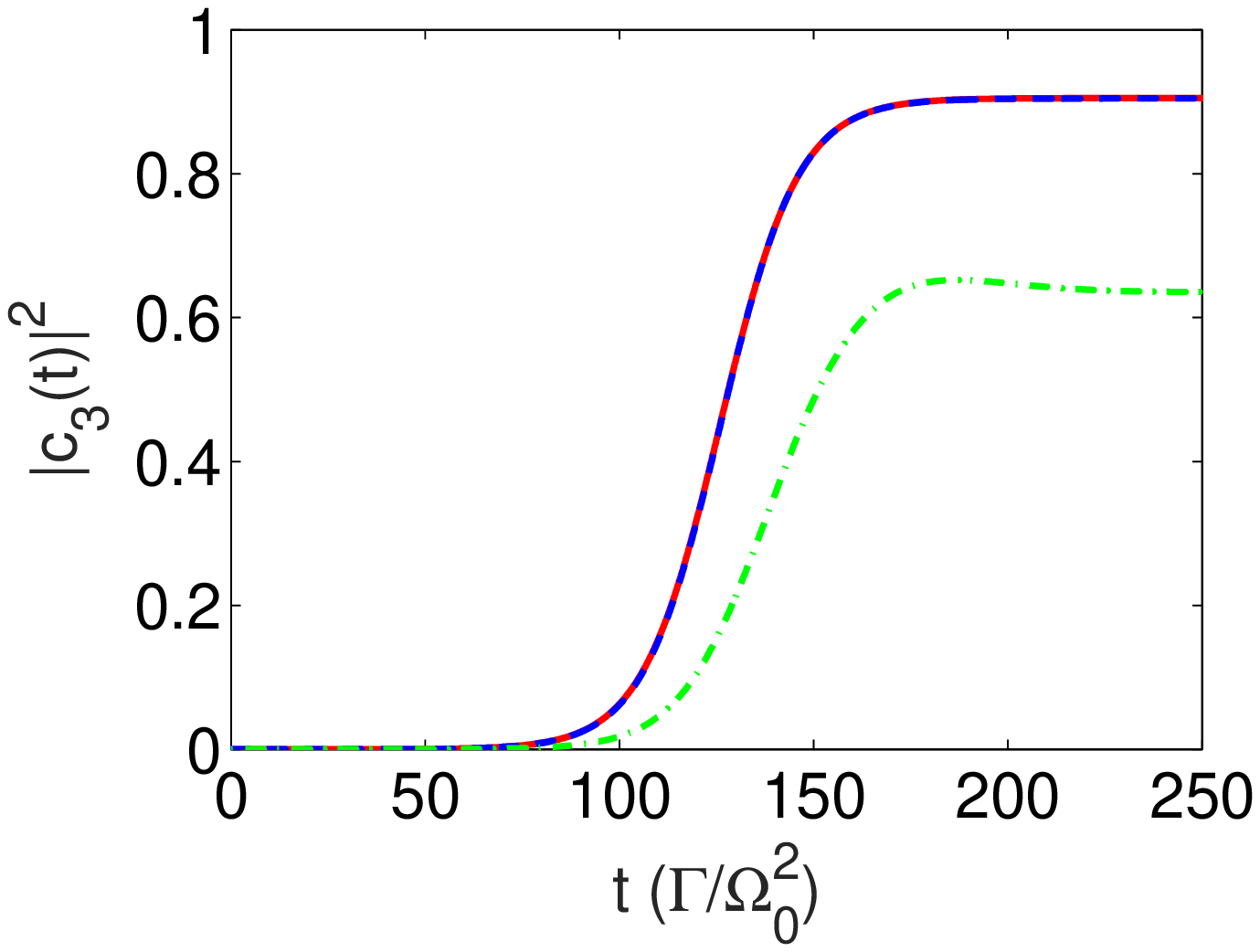}}
		\end{tabular}
\caption{(a) Time evolution of population $|c_3(t)|^2$ using the optimal pulse-sequence in the original system (\ref{STIRAP}) for $\Gamma\gg\Omega_0$ (red solid line), $\Gamma/\Omega_0=1$ (blue dashed line) and $\Gamma/\Omega_0=0.1$ (green dashed-dotted line). (b) Similar to (a) but using the adiabatic controls (\ref{adiabatic_controls}) in the original system.}
\label{fig:finite_gamma}
\end{figure}

Note that Fig. \ref{fig:example} has been obtained using the approximate system (\ref{eliminated}), which in principle is valid for $\Gamma\gg\Omega_0$. In order to test the validity of this approximation, in Fig. \ref{fig:opt_finite} we plot $|c_3(t)|^2$ using the optimal pulse-sequence in the original system (\ref{STIRAP}) for $\Gamma\gg\Omega_0$ (red solid line), $\Gamma/\Omega_0=1$ (blue dashed line) and $\Gamma/\Omega_0=0.1$ (green dashed-dotted line). Observe that the first two cases can be hardly distinguished and only the last one deviates, though remaining close in the efficiency for this large normalized duration $T=250$, as explained in the previous section. In Fig. \ref{fig:ad_finite} we display similar plots but using the adiabatic controls in the original system (\ref{STIRAP}). The cases with $\Gamma\gg\Omega_0$ (red solid line) and $\Gamma/\Omega_0=1$ (blue dashed line) almost coincide, as before, but the case with $\Gamma/\Omega_0=0.1$ (green dashed-dotted line) now deviates substantially. Note that for pulses (\ref{adiabatic_controls}) the derivative of the mixing angle is
\begin{equation}
\label{dot_th_ad}
\dot{\theta}=\frac{\Omega^2_0}{4\Gamma\bar{T}}\frac{1}{\cosh{\left(\frac{t-t_0}{2\bar{T}}\right)}},
\end{equation}
where recall that $\bar{T}$ is normalized. Using the maximum value of this $\dot{\theta}$ (for $t=t_0$) in the adiabaticity condition for the original system $\dot{\theta}\ll \Omega_0/2$ (neglecting dissipation), we obtain the condition
\begin{equation}
\label{adiabaticity_adiabatic}
\frac{1}{2\bar{T}}\ll \frac{\Gamma}{\Omega_0},
\end{equation}
which is not satisfied for $\bar{T}=10$ when $\Gamma/\Omega_0=0.1$.

\section{Conclusion}

In this work, we considered the problem of maximizing STIRAP efficiency in the limit where the dissipation rate of the intermediate state is much larger than the maximum amplitude of the Stokes and pump fields. Using this assumption we reduced the STIRAP system to a system involving only the initial and target states, and solved the optimal control problem of maximizing the population of the target state at a given final time, under the constraint that the sum of the intensities of the pump and Stokes pulses is constant, thus the only control parameter is actually the mixing angle of the fields. We found that in the optimal solution the mixing angle changes in a bang-singular-bang manner, where the initial and final bangs correspond to equal instantaneous rotations, while the intermediate part to a linear variation with time. The optimal angle corresponding to the initial and final rotations can be found by solving a transcendental equation where the duration of the process appears as a parameter, while the slope of the intermediate linear variation and the optimal transfer efficiency are expressed as functions of this angle. The counterintuitive pulse-sequence with nonzero fields at the boundary times is recovered as the optimal solution. We also demonstrate with numerical simulations that, transfer efficiency values close to the optimal derived using the approximate system, can also be obtained with the original STIRAP system using dissipation rates comparable to the maximum control amplitude.


%
%


\begin{thebibliography}{}
%
%

\bibitem{Bergmann98}
K. Bergmann, H. Theuer, and B. W. Shore, Coherent population transfer among quantum states of atoms and molecules, Rev. Mod. Phys., 70, 1003 (1998)

\bibitem{Kral07}
P. Kr\'{a}l, I. Thanopulos, and M. Shapiro, Coherently controlled adiabatic passage, Rev. Mod. Phys., 79, 53 (2007)

\bibitem{Vitanov17}
N. V. Vitanov, A. A. Rangelov, B. W. Shore, and K. Bergmann, Stimulated Raman adiabatic passage in physics, chemistry, and beyond, Rev. Mod. Phys., 89, 015006 (2017)

\bibitem{Sola18}
I. R. Sol\'{a}, B. Y. Chang, S. A. Malinovskaya and V. S. Malinovsky, Quantum control in multilevel systems, Adv. At. Mol. Opt. Phys., 67, 151-256 (2018)

\bibitem{Bergmann19}
K. Bergmann et al., Roadmap on STIRAP applications, J. Phys. B: At. Mol. Opt. Phys., 52, 202001 (2019)

\bibitem{Vitanov97}
N. V. Vitanov and S. Stenholm, Population transfer via a decaying state, Phys. Rev. A, 56, 1463-1471 (1997)

\bibitem{Kobrak98}
M. N. Kobrak and S. A. Rice, Coherent population transfer via a resonant intermediate state: The breakdown of adiabatic passage, Phys. Rev. A, 57, 1158-1163 (1998)

\bibitem{Paspalakis06}
E. Paspalakis, Adiabatic three-waveguide directional coupler, Opt. Commun., 258, 30-34 (2006).

\bibitem{Dreisow09}
F. Dreisow, A. Szameit, M. Heinrich M, R. Keil, S. Nolte, A. T\"{u}nnermann, and S. Longhi, Adiabatic transfer of light via a continuum in optical waveguides, Opt. Lett., 34, 2405-2407 (2009)

\bibitem{Menchon16}
R. Menchon-Enrich, A. Benseny, V. Ahufinger, A. D. Greentree, T. Busch and J. Mompart, Spatial adiabatic passage: a review of recent progress, Rep. Prog. Phys., 79, 074401 (2016)

\bibitem{Golter14}
D. A. Golter and H. L. Wang, Optically driven Rabi oscillations and adiabatic passage of single electron spins in diamond, Phys. Rev. Lett., 112, 116403 (2014)

\bibitem{Kumar16}
K. S. Kumar, A. Veps\"{a}l\"{a}inen A, S. Danilin, and G. S. Paraoanu, Stimulated Raman adiabatic passage in a three-level superconducting circuit, Nat. Commun., 7, 10628 (2016).



\bibitem{Bryson}
Jr. A. E. Bryson and Y.-C. Ho, Applied Optimal Control: Optimization, Estimation, and Control. Taylor \& Francis Group, New York (1975).

\bibitem{Band94}
Y. B. Band and O. Magnes, Is adiabatic passage population transfer a solution to an optimal control problem?, J. Chem. Phys., 101, 7528 (1994)

\bibitem{Sola99}
I. R. Sol\'{a}, V. S. Malinovsky, and D. J. Tannor, Optimal pulse sequences for population transfer in multilevel systems, Phys. Rev. A, 60, 3081 (1999)

\bibitem{Kis02}
Z. Kis and S. Stenholm, Optimal control approach for a degenerate STIRAP, J. Mod. Opt., 49, 111-124 (2002)

\bibitem{Kumar11}
P. Kumar, S. A. Malinovskaya and V. S. Malinovsky, Optimal control of population and coherence in three-level $\Lambda$ systems, J. Phys. B: At. Mol. Opt. Phys., 44, 154010, (2011)

\bibitem{Boscain02}
U. Boscain, G. Charlot, J.-P. Gauthier, S. Gu\'{e}rin, and H.-R. Jauslin, Optimal control in laser-induced population transfer for two- and three-level quantum systems, J. Math. Phys., 43, 2107 (2002)

\bibitem{Rat12}
H. Yuan, C. P. Koch, P. Salamon, and D. J. Tannor, Controllability on relaxation-free subspaces: On the relationship between adiabatic population transfer and optimal control, Phys. Rev. A, 85, 033417 (2012)

\bibitem{Khaneja03}
N. Khaneja, T. Reiss, B. Luy, S. J. Glaser, Optimal control of spin dynamics in the presence of relaxation, J. Magn. Reson., 162, 311-319 (2003)

\bibitem{Stefanatos04}
D. Stefanatos, N. Khaneja, S. J. Glaser, Optimal control of coupled spins in the presence of longitudinal and transverse relaxation, Phys. Rev. A, 69, 022319 (2004)

\bibitem{Stefanatos05}
D. Stefanatos and N. Khaneja, Semidefinite programming and reachable sets of dissipative bilinear control systems, Proceedings of the 44th IEEE Conference on Decision and Control, Seville, Spain, 2811-2816 (2005)

\bibitem{Assemat12}
E. Ass\'{e}mat and D. Sugny, Connection between optimal control theory and adiabatic-passage techniques in quantum systems, Phys. Rev. A, 86, 023406 (2012)

\bibitem{Dalessandro20}
D. D'Alessandro, B. A. Sheller, and Z. Zhu, Time-optimal control of quantum lambda systems in the KP configuration, J. Math. Phys., 61, 052107 (2020)


\bibitem{STA19}
D. Gu\'{e}ry-Odelin, A. Ruschhaupt, A. Kiely, E. Torrontegui, S. Mart\'{i}nez-Garaot, and J. G. Muga, Shortcuts to adiabaticity: Concepts, methods, and applications, Rev. Mod. Phys., 91, 045001 (2019)

\bibitem{Demirplak05}
M. Demirplak and S. A. Rice, Assisted adiabatic passage revisited, J. Phys. Chem. B, 109, 6838 (2005)

\bibitem{Chen10b}
X. Chen, I. Lizuain, A. Ruschhaupt, D. Gu\'{e}ry-Odelin, and J. G. Muga, Shortcut to adiabatic passage in two- and three-level atoms, Phys. Rev. Lett., 105, 123003 (2010)

\bibitem{Giannelli14}
L. Giannelli and E. Arimondo, Three-level superadiabatic quantum driving, Phys. Rev. A, 89, 033419 (2014)

\bibitem{Masuda15}
S. Masuda and S. A. Rice, Fast-forward assisted STIRAP, J. Phys. Chem. A, 119, 3479–3487, (2015)

\bibitem{Li16}
Y.-C. Li and Xi Chen, Shortcut to adiabatic population transfer in quantum three-level systems: Effective two-level problems and feasible counterdiabatic driving, Phys. Rev. A 94, 063411 (2016)

\bibitem{Clerk16}
A. Baksic, H. Ribeiro, and A.A. Clerk, Speeding up adiabatic quantum state transfer by using dressed states, Phys. Rev. Lett., 116, 230503 (2016)

\bibitem{Kolbl19}
J. K\"{o}lbl, A. Barfuss, M. S. Kasperczyk, L. Thiel, A. A. Clerk, H. Ribeiro, and P. Maletinsky, Initialization of single spin dressed states using shortcuts to sdiabaticity, Phys. Rev. Lett., 122, 090502 (2019)

\bibitem{Dridi20}
G. Dridi, K. Liu, and S. Gu\'{e}rin, Optimal robust quantum control by inverse geometric optimization, Phys. Rev. Lett., 125, 250403 (2020)

\bibitem{Petiziol20}
F. Petiziol, E. Arimondo, L. Giannelli, F. Mintert, and S. Wimberger, Optimized three-level quantum transfers based on frequency-modulated optical excitations, Sci. Rep., 10, 2185 (2020)




\bibitem{Stefanatos20f}
D. Stefanatos, A. Smponias, H. R. Hamedi, and E. Paspalakis, Ultimate conversion efficiency bound for the forward double-$\Lambda$ atom-light coupling scheme
Opt. Lett., 45, 6090-6093 (2020)

\bibitem{Lapert10}
M. Lapert, Y. Zhang, M. Braun, S. J. Glaser, and D. Sugny, Singular extremals for the time-optimal control of dissipative spin $1/2$ particles, Phys. Rev. Lett., 104, 083001 (2010)

\bibitem{Lin20}
C. Lin, D. Sels, and Y. Wang, Time-optimal control of a dissipative qubit, Phys. Rev. A, 101, 022320 (2020)

\bibitem{Stefanatos17a}
D. Stefanatos, Maximising optomechanical entanglement with optimal control, Quantum Sci. Technol., 2, 014003, (2017)

\bibitem{Paspalakis02}
E. Paspalakis and Z. Kis, Enhanced nonlinear generation in a three-level medium with spatially dependent coherence, Opt. Lett., 27, 1836-1838 (2002)

\bibitem{Hamedi19}
H.R. Hamedi, E. Paspalakis, G. \v{Z}labys, G. Juzeli\={u}nas, and J. Ruseckas, Complete energy conversion between light beams carrying orbital angular momentum using coherent population trapping for a coherently driven double-$\Lambda$ atom-light-coupling scheme Phys. Rev. A, 100, 023811 (2019); Erratum, 102, 019903(E) (2020)

\end{thebibliography}


\end{document}